\documentclass{aa}
\usepackage{dblfloatfix}
\usepackage{graphicx}
    \usepackage[font=small,labelfont=bf]{caption}
\usepackage{amsmath}
\usepackage{amssymb}
\usepackage{xfrac}
\usepackage{txfonts}

\usepackage{color}
\definecolor{Navy}		{RGB}{  0,   0, 128}
\definecolor{MidnightBlue}	{RGB}{ 25,  25, 112}
\definecolor{yellow}   	{RGB}{255, 215,   0}
\definecolor{darkorange}{RGB}{255, 140,   0}
\definecolor{dodgerblue}{RGB}{ 30, 144, 255}
\definecolor{black}     {RGB}{  0,   0,   0}
\definecolor{dimgray}   {RGB}{105, 105, 105}
\definecolor{gray}   {RGB}{128, 128, 128}

\usepackage{hyperref}
\hypersetup{colorlinks=true, linkcolor=Navy, citecolor=Navy, urlcolor=Navy}

\newcommand{\Espresso}{{Espresso}}

\newcommand\VAR[1]{#1}

\DeclareRobustCommand{\ion}[2]{%
  \relax
  \ifmmode
    \ifx\testbx\f@series
      {\mathbf{#1\,\mathsc{#2}}}
    \else
      {\mathrm{#1\,\mathsc{#2}}}
    \fi
  \else
    \textup{#1\,{\mdseries\textsc{#2}}}%
  \fi
 }

\begin{document}

\title{Chromatic Drift of the \Espresso{} Fabry-P\'erot Etalon}
\titlerunning{Chromatic Drift of the \Espresso{} FP Etalon}

\authorrunning{ T. M. Schmidt et al. }
          
\author{ Tobias M. Schmidt\inst{1}\and ... }
\author{ Tobias M. Schmidt\inst{1} \and Bruno Chazelas\inst{1} \and Christophe Lovis\inst{1} \and Xavier Dumusque\inst{1} \and Fran\c{c}ois Bouchy\inst{1} \and Francesco Pepe\inst{1} \and Pedro Figueira\inst{2,3} \and Danuta Sosnowska\inst{1} 
}

\institute{
    Observatoire Astronomique de l'Universit\'e de Gen\`eve, Chemin Pegasi 51, Sauverny, CH-1290, Switzerland \and
    European Southern Observatory, Alonso de C\'{o}rdova 3107, Vitacura, Región Metropolitana, Chile \and
    Instituto de Astrof\'{i}sica e Ci\^{e}ncias do Espa\c{c}o, Universidade do Porto, CAUP, Rua das Estrelas, 4150-762 Porto, Portugal
}

\date{\today}

\abstract{ 
In the last decade, white-light illuminated Fabry-P\'erot interferometers wave been established as a widely used, relatively simple, reliable, and cost-effective way to precisely calibrate high-resolution echelle spectrographs. 
However, \citet{Terrien2021} recently reported a chromatic drift of the Fabry-P\'erot interferometer installed at the Habitable-zone Planet Finder spectrograph. 
In particular, they found that the variation of the etalon effective gap size is not achromatic as usually assumed but in fact depends on wavelength.
Here, we present a similar study of the \Espresso{} Fabry-P\'erot interferometer. 
Using daily calibrations spanning a period of over 2.5~years, we also find clear evidence for a chromatic drift with an amplitude of a few cm/s per day that has a characteristic, quasi-oscillatory dependence on wavelength.
We conclude that this effect is probably caused by an aging of the dielectric mirror coatings and expect that similar chromatic drifts might affect all Fabry-P\'erot interferometers used for calibration of astronomical spectrographs.
However, we also demonstrate that the chromatic drift can be measured and in principle corrected using only standard calibrations based on hollow cathode lamp spectra.
}

\keywords{ Instrumentation: spectrographs -- Techniques: spectroscopic -- Techniques: radial velocities }

\maketitle

\section{Introduction}
\label{Sec:Introduction}

High-resolution echelle spectrographs are workhorses for astronomical research, however, many scientific application crucially rely on highly precise and accurate wavelength calibrations. Most famous in this context is the search for extrasolar planets using the radial-velocity method \citep[e.g.][]{Mayor1995} but also studies in the field of cosmology and fundamental physics \citep[e.g.][]{Sandage1965, Webb1999, Murphy2021} pose very demanding requirements on the spectrograph's wavelength calibration. This necessarily requires dedicated and specialized wavelength calibration sources.

In the last decade, white-light illuminated Fabry-P\'erot etalons (FP, \citealt{Perot1899}) have been established as relatively simple and cost-effective but nevertheless highly precise means of wavelength calibration for echelle spectrographs and are now widely used at many observatories \citep[see e.g.][]{Wildi2010, Wildi2011, Wildi2012, Bauer2015, Sturmer2017, Seifahrt2018, Hobson2021}.
They produce over a very broad spectral range a dense train of similarly bright lines which are (almost) equally spaced in frequency space, similar to a comb, and are therefore excellent wavelength calibration sources for high-resolution spectrographs.

While the width of the individual lines is defined by the reflectivity of the etalon surfaces, the spectral separation of the lines is given by the size of the gap in between the two etalon mirrors. Therefore, both parameters, linewidth and line separation, can rather easily be engineered to suit the needs of a given spectrograph. 
More specific, for a Fabry-P\'erot interferometer with physical separation $D$ of the etalon surfaces, the wavelength of a specific line with index $k$ is given by
\begin{equation}
 \lambda_k = \frac{2 \; D \; n }{k},
 \label{Eq:FP}
\end{equation}
where $n$ describes the index of refraction of the medium that fills the space in between the two mirrors.

Following this, the transmission spectrum of a Fabry-P\'erot etalon is subject to changes in the environmental conditions, namely temperature and pressure, which can have an influence on the physical spacing of the mirrors and the index of refraction. To limit this, the etalons are usually placed within a temperature and pressure controlled vacuum vessel and built from materials with very little thermal expansion. Still, the spectrum of such a passively stabilized Fabry-P\'erot interferometer is in principle \textit{freely floating}, in the sense that is not anchored to a wavelength reference, and will (slowly) drift with time.
Therefore, a Fabry-P\'erot interferometer does not provide any absolute wavelength information but always has to be used in conjunction with a second calibration source that does provide absolute wavelength information. In most cases, this is a thorium-argon or uranium-neon hollow cathode lamp. 
The sparse, with high dynamic range, and unevenly distributed atomic transitions of a hollow cathode lamp are by themselves completely insufficient to define a precise and accurate wavelength solution for a high-resolution echelle spectrograph%
\footnote{For example, \citet{Cersullo2019} showed that the thorium-only wavelength solution of the HARPS spectrograph suffers from deviations on intra-order scales of up to $200\,\mathrm{m/s}.$}%
, however, they provide enough information to accurately characterize the 
gap size of a Fabry-P\'erot interferometer, which then acts as a rather sophisticated method to precisely interpolate in between the atomic lines. The joint information from a Fabry-P\'erot interferometer and a hollow cathode lamp can therefore be utilized to establish a high fidelity wavelength solution and to precisely track the drift of stabilized spectrographs \citep{Bauer2015, Cersullo2019}.

Unfortunately, the FP gap size is in contrast to the simple form given in Equation~\ref{Eq:FP} not a constant but depends on wavelength. This is usually attributed to the use of dielectric coatings on the etalon mirrors which--simply speaking--cause light of different wavelength to be reflected at different depth \citep{Wildi2010, Bauer2015}. Therefore, Equation~\ref{Eq:FP} has to be modified by introducing a wavelength-dependent \textit{effective gap size} $D_\mathrm{eff}(\lambda)$ instead of just $D$ that describes the length of the Fabry-P\'erot cavity as function of wavelength. Nevertheless, $D_\mathrm{eff}(\lambda)$ is in practice a smooth and slowly varying function that can be accurately determined by comparing Fabry-P\'erot spectra to spectra of hollow cathode lamps, as described e.g. in \citet{Cersullo2019} or \citet{Schmidt2021}.

In addition, it is often assumed that the wavelength dependence of the Fabry-P\'erot gap size is independent from its time evolution, i.e. that there is some function $D_\mathrm{eff}(\lambda)$ that describes the spectral properties and is likely related to the structure of the dielectric coating but is by itself static and has in principle only to be measured once and in addition to that a drift $D(t)$ which has to be measured regularly (e.g. before and after each night) but affects all wavelengths in the same way by just increasing or decreasing the gap size by a constant value. The latter effect could be attributed to an expansion or shrinkage of the spacer separating the etalon mirrors, e.g. due to thermal variations, or a change of the residual pressure in the vacuum vessel. 
This concept has even lead to the development of actively stabilized Fabry-P\'erot interferometers \citep[see e.g.][]{Sturmer2017}, in which the cavity is at a single frequency locked against a stabilized laser, hoping that this scheme eliminates any drift at all wavelengths.

Recently, however, \citet{Terrien2021} presented a detailed analysis of the Fabry-P\'erot interferometer of the near-IR Habitable-zone Planet Finder spectrograph \citep[HPF,][]{Mahadevan2014} installed at the Hobby-Eberly Telescope at the McDonald Observatory in Texas. 
By comparing the spectrum of the Fabry-P\'erot interferometer to the fully independent and highly accurate wavelength solution derived from the laser frequency comb system at HPF \citep{Metcalf2019b, Metcalf2019a}, they were able to characterize the Fabry-P\'erot interferometer in detail and precisely track its drifts over a period of 6~months. Unsurprisingly, \citet{Terrien2021} found a global bulk velocity shift of $\mathrm{\approx -2\,cm/s\,/\,day}$. However, the measured drift depends quite strongly on wavelength, varies between $\mathrm{-7}$ and $\mathrm{+4\,cm/s\,/\,day}$, and shows a semi-periodic behavior as function of wavelength. This \textit{chromatic drift} of the Fabry-P\'erot interferometer, which actually dominates over the bulk velocity drift, was not expected and seen by \citet{Terrien2021} for the first time with an astronomical spectrograph.   

Here, we present a very similar study of the Fabry-P\'erot etalon of the  Echelle SPectrograph for Rocky Exoplanets and Stable Spectroscopic Observations (\Espresso{}, \citealt{Molaro2009, Pepe2010, Pepe2014, Pepe2020}) installed at the incoherently-combined Cound\'e focus of the ESO Very Large Telescope. 
\Espresso{} is in regular operation since November 2018. We therefore have daily calibration data available for a total period of \VAR{2.5\,years}, but with a gap of about eight months in Summer 2020 during which the observatory was closed.
This extensive dataset allows us to explore if also the \Espresso{} Fabry-P\'erot interferometer shows a similar chromatic drift as seen with the HPF spectrograph and determine whether the findings reported by \citet{Terrien2021} are a peculiar effect of the HPF system or in fact a general property that is--despite quite different designs--present in all astronomical Fabry-P\'erot interferometers.

\section{Data}
\label{Sec:Data}

For this study, we use all calibrations in \texttt{singleHR} mode with a binning of 1$\times$1 \citep[see][for details]{Pepe2020} taken between \VAR{first of November 2018} and \VAR{end of May 2021}. In this mode, \Espresso{} offers a spectral resolution of $R = \frac{\lambda}{\Delta\lambda} \approx 145\,000$ and is in 1$\times$1 binning used for the majority of bright radial-velocity targets. In addition, the calibration spectra taken in this mode have a signal-to-noise ratio (S/N) about $\sqrt{2}$ higher than for 2$\times$1 binning and provide the largest dataset, since calibrations for the ultra-high resolution mode, which would offer lower S/N but substantially higher resolution, are not taken on a daily basis. The instrument calibration plan foresees calibrations to be taken every morning, composed of bias frames, spectral flatfield exposures, as well as Fabry-P\'erot and thorium-argon spectra. Additional Fabry-P\'erot and thorium-argon exposures are taken in the late afternoons or evenings. Therefore, we usually have two full calibrations available per day.
The total dataset used in this study contains 1658 epochs and spans 941 days, however, there are no calibrations available between 2020-03-21 and 2020-11-29 due to a complete closure of the Paranal observatory and a shutdown of all operations.


The \Espresso{} Fabry-P\'erot device is based on the long legacy of very similar devices build at Geneva Observaory and follows the same design as described in \citet{Wildi2010, Wildi2011, Wildi2012}. It covers the full wavelength range of \Espresso{} from $3800\,\mathrm{\AA}$ to $7900\,\mathrm{\AA}$ and is designed with a mirror spacing of approximately  $7.1\,\mathrm{mm}$, corresponding to a separation between modes of slightly less than $20\,\mathrm{GHz}$. 
The nominal finesse is relatively low, approximately  $\mathcal{F}\approx12$, and the Fabry-P\'erot lines are noticeably resolved by the spectrograph. 
The Fabry-P\'erot device is kept in a vacuum vessel which is continuously pumped to keep the residual pressure at a level of approximately $10^{-5}\,\mathrm{mBar}$ and actively stabilized to a temperature slightly above ambient.

In principle, \Espresso{} is equipped with a Laser Frequency Comb \citep[LFC,][]{Megevand2014, Probst2014, Probst2016, Schmidt2021} that provides a highly precise and accurate wavelength solution and would be the ideal reference to characterize the Fabry-P\'erot interferometer. However, the \Espresso{} LFC still suffers from substantial reliability issues. For the given time period, LFC exposures are available in the ESO archive for only 124~days, distributed over the months of November 2018 and July to November 2019. In addition, the LFC offers only a limited wavelength coverage, corresponding to approximately 57\% of the total \Espresso{} spectral range.  
Therefore, we decided to rely for this study exclusively on the spectra of the thorium-argon (ThAr) hollow cathode lamp as absolute wavelength reference to characterize the Fabry-P\'erot interferometer. 
Clearly, the thorium-argon spectra have disadvantages, namely the much lower number, unequal spacing, and in general lower S/N of the lines. Also, there is some uncertainty to which degree the spectral lines are stable over long timescales or possibly affected by the aging of the lamp \citep[see e.g.][]{Nave2018}.
However, the much larger coverage in terms of spectral range and time provided by the thorium-argon spectra is essential for the purpose of this study and clearly outweighs the disadvantages. 
In the following, we show that, despite their imperfections, the thorium-argon spectra are sufficient to facilitate a detailed characterization of the Fabry-P\'erot interferometer and to draw solid conclusions about the presence of a chromatic drift.

All data was reduced with the \Espresso{} Data Reduction System \citep[DRS, version 2.3.1\,\footnote{\url{https://www.eso.org/sci/software/pipelines/index.html}},][]{Lovis2020} in the standard way as it is used for the Data \& Analysis Center for Exoplanets (DACE\,\footnote{\url{https://dace.unige.ch/}}) hosted at the Geneva Observatory.
We use the default ThAr line list incorporated in the \Espresso{} DRS \citep{Lovis2020}. This contains a relatively sparse selection of 407 unique thorium lines which were carefully selected to be unblended and of high quality. 
For various reasons, these lines appear multiple times in a single exposure, e.g. because \Espresso{} uses a pupil slicer which images each spectral order into two independent traces, but also due to substantial overlap of the individual echelle orders. In addition, we use calibration data from both optical fibers, since they provide equal and consistent information. Following this, we have for each epoch 2300 thorium line measurements available.

By comparing Fabry-P\'erot to thorium-argon spectra taken immediately after each other, one can derive from each individual thorium line position a measurement of the Fabry-P\'erot effective gap $D_\mathrm{eff}(t,\lambda)$ at the given time $t$ and wavelength $\lambda$. The concept behind this is explained in detail e.g. in \citet{Schmidt2021}. 
Briefly, the measured position of a thorium line is compared to the Fabry-P\'erot lines falling in the same region and from the known absolute wavelength of the spectral line one can, after assigning the correct indices to the Fabry-P\'erot lines, deduce the effective gap size.
If one would like to derive the combined ThAr/FP wavelength solution, one would have to construct a model for the effective gap from the individual $D_\mathrm{eff}$ measurements and then evaluate this model to get the wavelengths at every position on the detector. However, for the purpose of this study, this is not necessary. 
We simply use the $D_\mathrm{eff}(t,\lambda)$ measurements provided by the individual thorium lines and analyze their temporal behavior. Therefore, we do not have to invoke any specific model for the Fabry-P\'erot effective gap size and can treat all measurements to be formally independent.

\section{Chromatic FP Drift}

\begin{figure*}[htb]
 \centering
  \includegraphics[width=\linewidth]{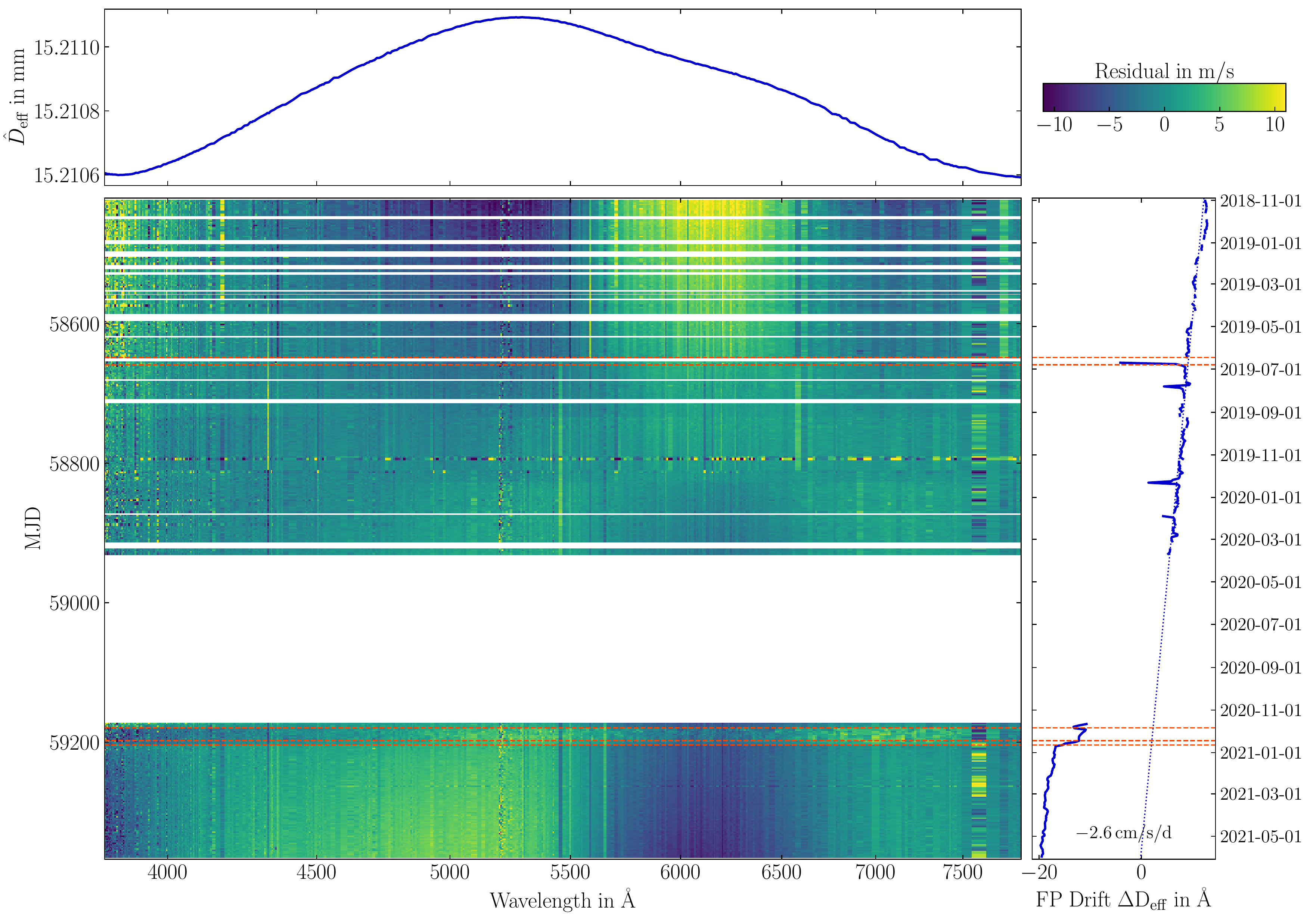}
  \caption{
   Characterization of the \Espresso{} Fabry-P\'erot effective gap size as function of time and wavelength, decomposed as described in Equation~\ref{Eq:Decomposition}.
   The top panel shows just the wavelength dependence of the cavity length $\hat{D}_\mathrm{eff}(\lambda)$ obtained by time-averaging all measurements.
   The right panel shows the achromatic drift $\Delta{D}_\mathrm{eff}(t)$ as function of time. A linear trend was fitted for the period before the shutdown in March 2020.
   The central panel shows the residual variation of the ${D}_\mathrm{eff}(t,\lambda)$ measurements after subtraction the average spectral shape and the achromatic drift. The values are expressed as velocity shift. Horizontal red lines correspond to interventions to the system.
   }
  \label{Fig:FP_Drift}
\end{figure*}

The dataset described above contains nearly four million $D_\mathrm{eff}(t,\lambda)$ measurements. To make the analysis and visualization easier, we bin the data in wavelength and time. For this, we combine all measurements that correspond to the same thorium line, which results in 407 wavelength bins of unequal size. In addition, we combine data in time to blocks of two days length. We apply uniform weights to the $D_\mathrm{eff}$ measurements contributing to a single bin.

The largest variation in the $D_\mathrm{eff}(t,\lambda)$ measurements stems from the wavelength-dependent change of the Fabry-P\'erot cavity length and is much larger than the expected chromatic drift. Thus, we decompose the $D_\mathrm{eff}(t,\lambda)$ measurements according to the concepts presented in Section~\ref{Sec:Introduction} into the following terms:
\begin{equation}
 D_\mathrm{eff}(t,\lambda) = \hat{D}_\mathrm{eff}(\lambda) + \Delta{}D_\mathrm{eff}(t) + R(t,\lambda).
 \label{Eq:Decomposition}
\end{equation}
Here, $\hat{D}_\mathrm{eff}(\lambda)$ represents the mean effective cavity length as function of wavelength and is obtained by averaging all measurements over time
\begin{equation}
 \hat{D}_\mathrm{eff}(\lambda) = \left\langle \; {D}_\mathrm{eff}(t,\lambda) \; \right\rangle_{t}.
\end{equation}
The temporal but achromatic drift is described by $\Delta{}D_\mathrm{eff}(t)$ which is computed by first subtracting the mean spectral dependence from the datapoints and then averaging over wavelength
\begin{equation}
 \Delta{D}_\mathrm{eff}(t) = \left\langle \; {D}_\mathrm{eff}(t,\lambda) - \hat{D}_\mathrm{eff}(\lambda) \; \right\rangle_{\lambda}.
\end{equation}
The remaining term $R(t,\lambda)$ contains all residuals that are not captured by the wavelength- and time-dependent variations alone.
Therefore, it is representative of the chromatic component of the Fabry-P\'erot drift.

Our characterization of the \Espresso{} Fabry-P\'erot is visualized in Figure~\ref{Fig:FP_Drift}.
The top panel shows the average effective gap size $\hat{D}_\mathrm{eff}(\lambda)$ as function of wavelength. Here, we have absorbed the factor 2 and the index of refraction%
\footnote{Although the Fabry-P\'erot device is located in a vacuum vessel small changes of the residual pressure and therefore the index of refraction have a measurable effect on the output spectrum.} 
given in Equation~\ref{Eq:FP} as well as some potential geometric factor in the $\hat{D}_\mathrm{eff}(\lambda)$ quantity%
\footnote{Some geometric corrections factor will stem from the fact that the \Espresso{} FP is fed by a multimode fiber which has a finite extension compared to the theoretical assumption of a perfectly collimated beam produced by a point source, see e.g. \citet{Hao2021}.}%
. The geometric spacing of the \Espresso{} etalon is therefore only half of the shown $D_\mathrm{eff}\approx15.210\,\mathrm{mm}$ and the wavelengths of individual Fabry-P\'erot modes are simply $\lambda_k = \frac{ D_\mathrm{eff} }{ k }$.

The quantity shown in the top panel of Figure~\ref{Fig:FP_Drift} is in principle equivalent to the $D_\mathrm{eff}(\lambda)$ presented in Figure~11 of \citet{Schmidt2021}. 
The apparent discrepancy results from a different choice for the Fabry-P\'erot line indices.
The measured $D_\mathrm{eff}$ value depends on the index one assigns to a given line. However, the absolute value of this index is not uniquely defined, in particular not when the effective gap varies substantially with wavelength, and in consequence also the value of $D_\mathrm{eff}$ is ambiguous.
For the studies presented here and in \citet{Schmidt2021} different data reduction codes were used which implement different choices for the line index, resulting in a global offset in $k$ and therefore a slightly different value of $D_\mathrm{eff}$. However, this is of no relevance in practice as long as one always uses a consistent combination of indexing and effective gap description \citep[see e.g.][]{Bauer2015}. 

\begin{figure*}[htb]
 \centering
 \includegraphics[width=\linewidth]{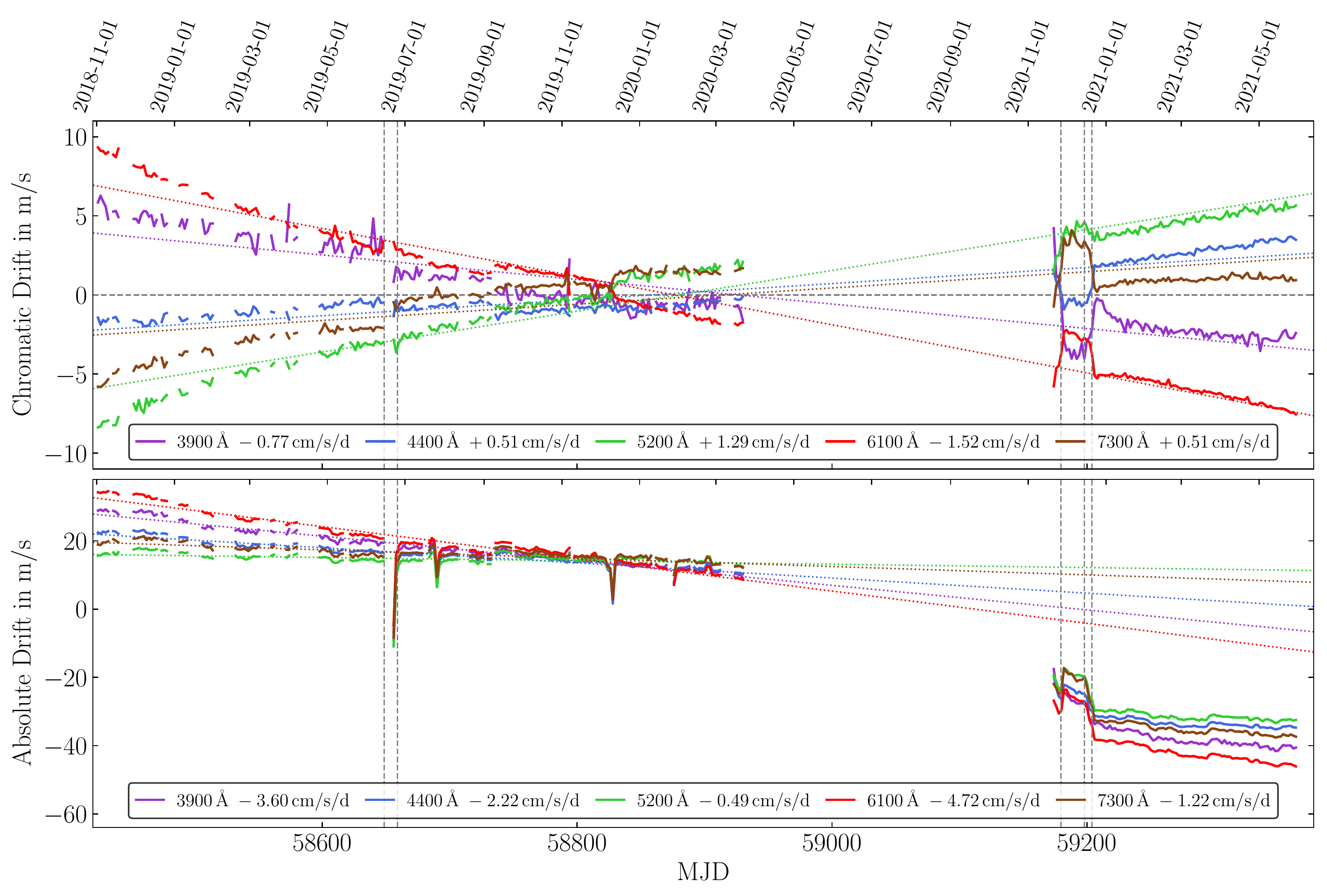}
 \caption{Chromatic drift of the \Espresso{} Fabry-P\'erot for five representative wavelengths. The curves in the top panel show averages of the measurements displayed in the central panel of Figure~\ref{Fig:FP_Drift} obtained in $\pm4\%$ wide spectral ranges around the stated wavelengths. Linear slopes were fitted to the data to provide a coarse estimate of the chromatic drift rates. The bottom panel shows basically the same data, however, without subtracting the achromatic drift. Linear slopes were only fitted to epochs before the shutdown.
  }
 \label{Fig:FP_DriftRates}
\end{figure*}

The right panel of Figure~\ref{Fig:FP_Drift} shows the achromatic component of the drift $\Delta{D}_\mathrm{eff}(t)$ as a function of time.
Most obvious is the 8~month long gap of this time series in Summer 2020 caused by the closure of the observatory. After the ramp-up of \Espresso{} in November~2020, the Fabry-P\'erot interferometer did not return to the same state it was before the shutdown but settled to an effective gap that is $\approx 20\,\mathrm{\AA}$ smaller%
\footnote{Also the spectrograph did not return to exactly the same configuration which manifests itself e.g. by a shift of the echellogram by several pixels.}.
We fitted a linear function to the $\Delta{D}_\mathrm{eff}(t)$ data obtained before the shutdown and determined a drift of $-2.5\,\mathrm{cm/s\,/\,d}$. In 2021, this drift seems to continue, although with the stated offset.
The right panel of Figure~\ref{Fig:FP_Drift} also shows several excursion from this general trend. Some of these correspond to known interventions to the instrument, indicated by horizontal red lines. Most notably, the full \Espresso{} fiber feed was replaced in July 2019 and there were several changes of the light source illuminating the Fabry-P\'erot interferometer in November and December 2020. The sharp spike in December 2019 corresponds to an anomaly of temperature and pressure in the Fabry-P\'erot vacuum vessel.

The central panel of Figure~\ref{Fig:FP_Drift} shows the residuals $R(t,\lambda)$ obtained by subtracting the mean spectral and mean temporal behavior from the initial $D_\mathrm{eff}(t,\lambda)$ measurements. This residual variation of the Fabry-P\'erot gap size has been converted to a velocity, the quantity commonly used within the context of radial-velocity studies.
As stated above, there is a significant offset between the measurements taken before and after the shutdown, however, subtracting the achromatic drift captures basically all of this, allowing to visualize and analyze the residuals for the whole available period without having to deal with any strong discontinuity.
As described in Section~\ref{Sec:Data}, our characterization of the Fabry-P\'erot is based on the thorium lines and not all of them are perfectly well-behaved. This leads to some vertical striping in the central panel of Figure~\ref{Fig:FP_Drift} and to a few columns with particular high noise, predominantly towards the ends of the spectral range. However, despite these systematics, the global trend and in particular a clear chromatic drift of the \Espresso{} Fabry-P\'erot is easily visible%
. %
The most-noticeable features in Figure~\ref{Fig:FP_Drift} are large-scale coherent structures, corresponding to a drift that depends on wavelength. 
This can e.g. be seen around $6000\,\mathrm{\AA}$ where we find a residual that evolves from $+10\,\mathrm{m/s}$ to $-9\,\mathrm{m/s}$ over the course of 2.5~years.  At $\approx5000\,\mathrm{\AA}$, however, we find a drift with the opposite sign evolving from $-9\,\mathrm{m/s}$ to $+8\,\mathrm{m/s}$. At wavelengths shorter than $4200\,\mathrm{\AA}$ this becomes a negative drift again.

In the top panel of Figure~\ref{Fig:FP_DriftRates}, we show the residual drift vs. time for a few representative wavelength. Since the achromatic drift $\Delta{D}_\mathrm{eff}(t)$ is already subtracted, any significant deviation from zero is a clear indication for a chromatic drift.
The drifts at the different wavelengths clearly do not follow a linear relation and the absolute value of the drift rates seems to be larger at earlier times. Nevertheless, we fit straight lines to the data to get an approximate estimate of the magnitude of the chromatic drift. We find values between $-1.7\,\mathrm{cm/s\,/\,d}$ at $6100\,\mathrm{\AA}$ and $+1.25\,\mathrm{cm/s\,/\,d}$ at $5200\,\mathrm{\AA}$.
However, there is no simple, e.g. monotonic relation between wavelength and drift rates. 
The bottom panel of Figure~\ref{Fig:FP_DriftRates} shows basically the same, but in terms of absolute drift rates, i.e. without subtracting the achromatic drift.

\begin{figure*}[htb]
 \centering
 \includegraphics[width=\linewidth]{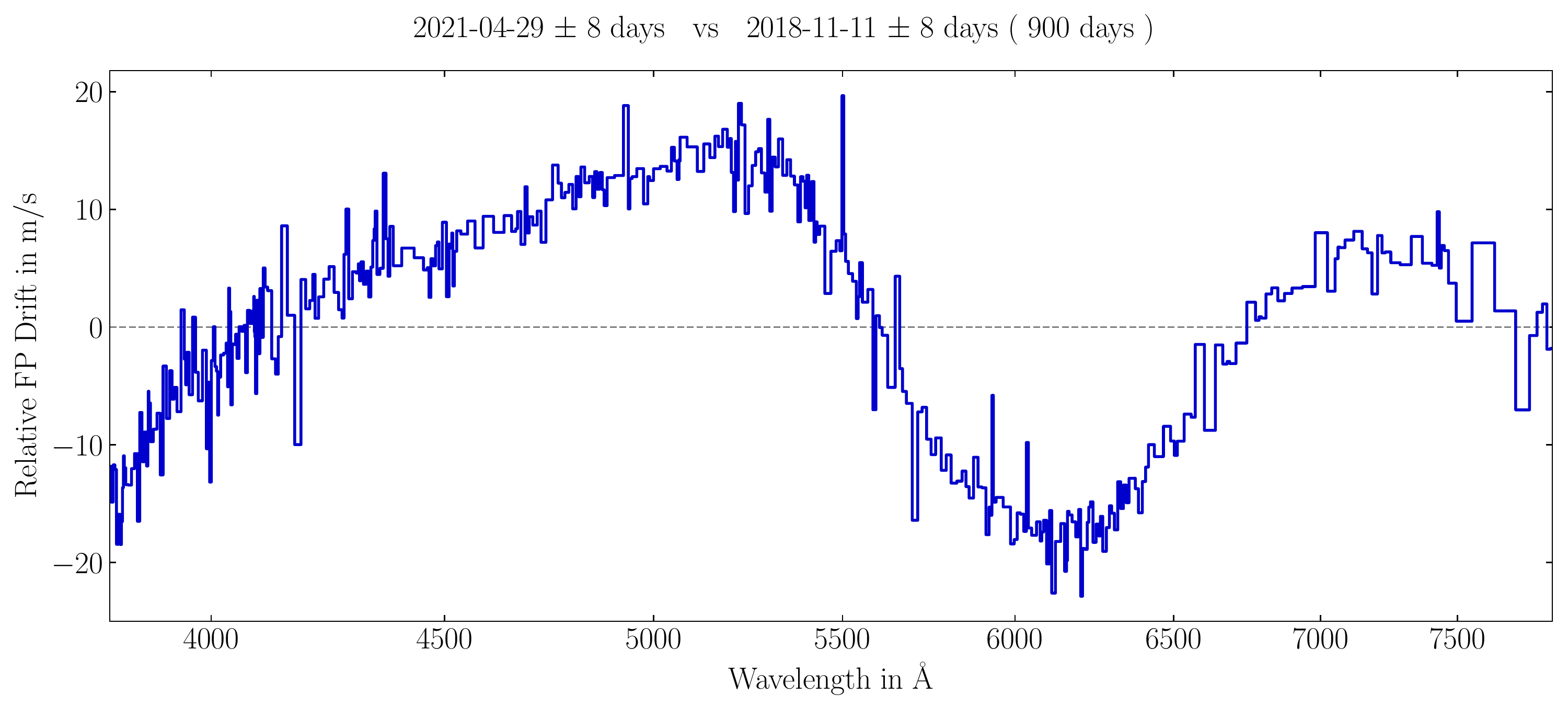}
 \caption{
  Wavelength dependence of the chromatic drift of the \Espresso{} Fabry-P\'erot etalon over a period of 900~days. The achromatic drift $\Delta{}D_\mathrm{eff}(t)$ has already been subtracted.
  }
 \label{Fig:FP_DriftWave_900}
\end{figure*}

To further visualize the dependence on wavelength, we show in Figure~\ref{Fig:FP_DriftWave_900} the difference of the effective gap size between the beginning and end of our timeseries as function of wavelength. For this, we combine all $R(t,\lambda)$ measurements from two 16~day long time periods (2021-04-29 $\pm8\,\mathrm{days}$ and 2018-11-11 $\pm8\,\mathrm{days}$) and compute the difference over this 900~day timebase. 
Figure~\ref{Fig:FP_DriftWave_900} clearly shows a quasi-oscillatory pattern for the chromatic drift as function of wavelength. The pattern is not strictly periodic but exhibits some wave-like pattern. The variation, however, is slow and we see over nearly a full octave in wavelength (or frequency) just a bit more than one period.

\section{Interpretation of the Data}

In principle, the observed drift could be caused by effects unrelated to the Fabry-P\'erot interferometer.
There are other aspects in the setup, like the spectrograph itself or the illumination of the etalon, that might as well have an impact on the observed line positions and therefore the inferred $D_\mathrm{eff}(t,\lambda)$.    
However, we are convinced that these do not dominate our measurement and that the observed chromatic drift is indeed caused by the Fabry-P\'erot interferometer. In the following, we discuss several possibilities and explain why these parasitic effects can not be the dominant source for the observed chromatic drift.

The best argument to rule out the spectrograph as a source for the observed drift is the absence of any discontinuity around $5200\,\mathrm{\AA}$. The \Espresso{} spectrograph is composed of two arms, both equipped with independent transfer collimators, cross-dispersers, cameras, and detectors \citep[see][]{Pepe2020}. The dichroic splits the light around $5200\,\mathrm{\AA}$. Still, Figure~\ref{Fig:FP_Drift} shows only slightly higher noise but no discontinuity in this region and the large-scale pattern of the chromatic drift is fully consistent and continuous across both arms. Therefore, a slow change of the spectrograph optical properties seems unlikely.
Also, the Fabry-P\'erot and thorium-argon spectra are taken immediately after each other and there is no reason why it should exhibit the smooth evolution with time. Therefore, a relative drift of the spectrograph can be excluded as well.  
In principle, a slow change of the spectrograph's line-spread function could induce an apparent drift between Fabry-P\'erot and thorium-argon spectrum since the intrinsic linewidth of both sources is substantially different. However, it is not plausible that such an effect causes the observed pattern and in particular not that it happens in the same way for both arms. 
In addition, we checked whether the \textit{beat pattern noise} described in \citet{Schmidt2021} could have a relevant impact. It was determined that this systematic displacement of spectral lines remains extremely stable over timescales of months and years and does not influence the measurement of the chromatic drift in a significant way. Also, we find no evidence for possible parasitic Fabry-P\'erot effects.
Furthermore, we confirm that both fibers of the spectrograph (science and sky fiber) deliver consistent results for the drift measurement.

Also, we exclude a possible shift of the thorium lines utilized for characterization of the Fabry-P\'erot interferometer as possible source of the observed chromatic drift.
Most of all, \citet{Nave2018} report that the wavelengths of the thorium lines are robust against changes of the pressure inside the lamp and therefor should not exhibit shifts due to aging of the hollow cathode lamp.
Even if this would be the case, the properties of the observed Fabry-P\'erot drift are not consistent with a possible shift of the thorium lines.
This can be seen in Figure~\ref{Fig:FP_DriftWave_900}. Each step in the shown function corresponds to the drift derived from one individual thorium line. Clearly, there are some outliers. However, it seems highly unlikely that the coherent, large-scale quasi-oscillatory pattern could be caused by a intrinsic shift of the thorium lines. 

Another reason for an apparent chromatic drift that is not directly related to the Fabry-P\'erot etalon could be a changing spectral energy distribution (SED) of the illuminating light source. A slope in the continuum illuminating the Fabry-P\'erot interferometer can lead to an offset in the fitted line centroid. With increasing age of the lightsource, this might change and induce an apparent drift.
We analyze the evolution of the Fabry-P\'erot spectral energy distribution and indeed find significant changes over the 2.5~year period. 
However, these do not correlate with the observed pattern of the chromatic drift.
Also, we determine that the slopes might introduce shifts of up to $10\,\mathrm{cm/s}$ but these remain stable at the few cm/s level over most of the 2.5~year period and are therefore orders of magnitude smaller than the observed chromatic drift.

A rather noticeable feature, however, appears in November and December 2020 with rapid, chromatic drifts up to $5\,\mathrm{m/s}$ (Figure~\ref{Fig:FP_Drift} and \ref{Fig:FP_DriftRates}). This is related to several changes of the lamp illuminating the Fabry-P\'erot interferometer. On November 27, the existing lightsource, an Energetiq laser-driven light source EQ-99X, which did show some wear at this point, was replaced by a supercontinuum source that in general should exhibit less degradation over time. However, this change of lamps had a negative impact on the quality of the wavelength calibration. In particular the \textit{beat pattern noise} \citep{Schmidt2021} increased substantially by about 30\,\%.
The new lightsource had a different spectral energy distribution than the old one and is coupled to a monomode instead of a multimode optical fiber, but 
there is no fully plausible explanation why this would cause the issues encountered in the wavelength calibration.
In any case, the problem was solved by switching back to the original lamp on December 17. Another lamp change happened on the 23rd of December when the exiting lamp was replaced by a newer spare item of the same type. 
These changes to the Fabry-P\'erot lightsource did have an influence on the measured chromatic drift and rather strong deviations from the overall pattern are visible in Figure~\ref{Fig:FP_Drift} and \ref{Fig:FP_DriftRates} around these dates. 
We confirm that the different spectral energy distribution of the lightsources can not be exclusively responsible for the chromatic drifts observed. Therefore, further effects must be in play which, however, could not be determined in detail.   
We have to conclude that the spectrum of the Fabry-P\'erot device cannot be assumed to be stable when the lightsource and therefore the illumination of the etalon is changed. It therefore seems advisable to limit intervention and modifications of these parts of the setup as much as possible. 

Apart from the events associated with the change of the lightsources in November and December 2020, we can exclude unrelated effects and conclude that with high certainty the observed chromatic drift stems indeed from the Fabry-P\'erot etalon itself. The most-plausible explanation for this is an aging of the dielectric mirror coatings. These are the only component in the setup that can introduce a strong chromatic effect.

Unfortunately, the exact layer properties of the dielectric stack for the \Espresso{} etalon mirror coatings are not available. Therefore, it is not possible to perform a quantitative modeling and determine how aging affects the measured effective gap size. 
In general, we can imagine several ways by which the coatings might evolve with time.
One possibility is that during manufacturing water vapor was trapped in the dielectric layers. The \Espresso{} Fabry-P\'erot uses \textit{soft coating} which are somewhat porous and known for trapping volatiles.  
Under vacuum conditions, the trapped water would evaporate, lead to a change of the index of refraction, and therefore to a change of the etalon effective spacing.
Figure~\ref{Fig:FP_DriftRates} indeed shows the fastest drift rates at the beginning of regular operations in 2018, but this was already about one year after installment of \Espresso{} at Paranal.  Figure~\ref{Fig:FP_DriftRates} also shows that even after years of operation under vacuum conditions, the drift continues with no asymptotic state in sight. It seems not very likely that an outgassing process as described above would persist for such long times.
Therefore, other processes are probably involved as well. With the data at hand, it is not possible to determine the root cause of the observed chromatic drift and further studies are required.

\section{Comparison with the HPF Faby-P\'erot}
\label{Sec:ComparisonTerrien}

Chromatic drift of a Fabry-P\'erot interferometer used as calibration source for an astronomical spectrograph was first reported by \citet{Terrien2021}. 
Here, we want to briefly compare the results of our analysis of the \Espresso{} Fabry-P\'erot interferometer with the ones presented by \citet{Terrien2021} of the one installed at the HPF spectrograph.

It has to be stressed that both instruments have several important differences.
The HPF spectrographs and its corresponding Fabry-P\'erot interferometer operates in the near infrared between $8200\,\mathrm{\AA}$ and $12800\,\mathrm{\AA}$, roughly covering the $z$, $y$, and $J$ bands, while \Espresso{} is sensitive in the visible domain between $3800\,\mathrm{\AA}$ and $7900\,\mathrm{\AA}$.
More importantly, the HPF Fabry-P\'erot etalon is designed with a high finesse of $\mathcal{F}\approx40$ to achieve a small intrinsic linewidth ($\approx800\,\mathrm{MHz}$ or 
$\approx0.85\,\mathrm{km/s}$), substantially narrower than the instrumental resolution \citep{Jennings2020}.
The \Espresso{} Fabry-P\'erot etalon, however, has a lower finesse of $\mathcal{F}\approx12$ and the Fabry-P\'erot lines are clearly resolved and appear about 20\,\% wider than the intrinsic linewidth.
Therefore, the mirrors in the HPF etalon have a much higher reflectivity ($\approx 90\%$) than the \Espresso{} ones ($\approx 76\%$).
In addition, the HPF etalon is fed by monomode fibers from a NKT SuperK supercontinuum lightsource while the \Espresso{} one uses multimode fibers and is illuminated by an Energetiq laser-driven light source EQ-99X. 
Therefore, the \Espresso{} and HPF Fabry-P\'erot interferometers do follow the same underlying idea but exhibit quite substantial differences in how this basic concept is implemented in practice. 

Nevertheless, we find overall quite similar results as \citet{Terrien2021}.  
They report an achromatic bulk drift of their Fabry-P\'erot interferometer of $\approx-2\,\mathrm{cm/s\,/\,d}$ while we find for \Espresso{} $-2.6\,\mathrm{cm/s\,/\,d}$ (see Figure~\ref{Fig:FP_Drift}).
Also the amplitude of the chromatic drift is similar. \citet{Terrien2021} report, depending on the wavelength, drifts between $-7\,\mathrm{cm/s\,/\,d}$ and $+3\,\mathrm{cm/s\,/\,d}$ when not subtracting the bulk drift. We find values between $-2.2\,\mathrm{cm/s\,/\,d}$ and $+1.6\,\mathrm{cm/s\,/\,d}$  after correcting for the achromatic drift (see Figure~\ref{Fig:FP_DriftRates} and \ref{Fig:FP_DriftWave_900} %
\footnote{Some apparent discrepancies in the drift rates between Figure~\ref{Fig:FP_DriftRates} and Figure~\ref{Fig:FP_DriftWave_900} result in the fact that the drift is not fully linear.}%
) and equivalently absolute drift rates between $-5.6\,\mathrm{cm/s\,/\,d}$ and $-0.2\,\mathrm{cm/s\,/\,d}$ determined over a 450~day period before the shutdown.    
Although the \Espresso{} chromatic drift is a bit smaller, this is a striking resemblance, given the rather different designs of the two Fabry-P\'erot devices.

Most informative is probably a comparison of the drift rate as function of wavelength. This is presented in Figure~7 in \citet{Terrien2021} and here in Figure~\ref{Fig:FP_DriftWave_900}. The HPF etalon shows a clear oscillatory pattern with a period that significantly increases with wavelength. Over the presented wavelength range from $8200\,\mathrm{\AA}$ to $12800\,\mathrm{\AA}$, six oscillations can be seen.
For \Espresso{}, the situation is less clear and we just find a quasi-oscillatory pattern. Despite the larger (relative) spectral range covered by our study, Figure~\ref{Fig:FP_DriftWave_900} only shows one clear oscillation and some aperiodic behavior bluewards of $5000\,\mathrm{\AA}$. 
In this sense, the two Fabry-P\'erot etalons show quite different behaviors.
On the other hand, one has to consider that the dielectric coatings on both of them have to be substantially different, achieving quite different reflectivity over different spectral ranges. 
Therefore, it is worth to highlight the similarity, i.e. the appearance of a (quasi-)oscillatory pattern, rather than the differences in detail.

Based on these similarities, we think that we basically observe the same effect for both Fabry-P\'erot interferometers and that the chromatic drift is related to an aging of the dielectric coatings. Also, it appears that chromatic drifts are probably a very common feature of Fabry-P\'erot interferometers that quite likely affects all such devices used in astronomy.    
The magnitude of the effect, i.e. a few cm/s per day, is not dramatic which explains why it could be easily overlooked in previous studies. Nevertheless, it will accumulate over longer periods. An uncorrected chromatic drift of the Fabry-P\'erot interferometer will cause radial velocities from different spectral ranges to become inconsistent with each other. 
{\color{black}Tests on standard stars reveal that this effect can actually be observed on-sky. Radial velocities extracted from  different spectral regions do indeed show a differential drift that is broadly consistent with the inverse of the Fabry-P\'erot chromatic drift we present in this study. }
It still has to be determined to which degree additional effects like variable seeing and wavelength-dependent fiber injection losses then lead to an overall impact on the measured radial velocity.
Nevertheless, this highlights the importance to correct for the chromatic drift of the Fabry-P\'erot interferometer when aiming for extreme-precision radial-velocity measurements.

Fortunately, this should be possible without major difficulties. In this study, we clearly demonstrate that the chromatic drift of the \Espresso{} Fabry-P\'erot interferometer can be adequately measured and characterized using the standard thorium-argon exposures. In addition, we show that the chromatic drift is well-behaved in the sense that it mostly affects the larges spectral and temporal scales. Therefore, it seems feasible to measure and correct for the chromatic drift as part of the routine wavelength calibration process without requiring any special calibration exposures.
{\color{black}Efforts to implement such a correction for the \Espresso{} Data Reduction System are underway.}

\section{Conclusions}

In this study we present a systematic analysis of the \Espresso{} Fabry-P\'erot interferometer. We use calibration data from 1658 epochs spread over 2.5~years from November 2018 to May 2021, but excluding eight months in Summer 2020.
Our analysis is based on the absolute wavelength information provided by the standard thorium-argon calibration spectra which deliver 2300 measurements of the Fabry-P\'erot effective gap size per epoch, derived from 407 unique thorium lines. 

We decompose these nearly four million $D_\mathrm{eff}(t,\lambda)$ measurements in a purely wavelength-dependent variation of the effective gap size $\hat{D}_\mathrm{eff}(\lambda)$, an achromatic time-evolution $\Delta{}D_\mathrm{eff}(t)$, and a residual component $R(t,\lambda)$ that reflects the chromatic drift of the Fabry-P\'erot interferometer. 
We find in these residuals clear, large-scale structure with an amplitude of up to $10\,\mathrm{m/s}$  (see Figure~\ref{Fig:FP_Drift}). There is no simple, i.e. monotonic, relation between wavelength and chromatic drift, instead we find positive as well as negative drift rates with an amplitude of up to $2.2\,\mathrm{cm/s\,/\,d}$, which is comparable with the achromatic drift of $-2.6\,\mathrm{cm/s\,/\,d}$ (Figure~\ref{Fig:FP_DriftRates}).
As a function of wavelength, the chromatic drift of the \Espresso{} Fabry-P\'erot etalon exhibits a wave-like, quasi-oscillatory pattern, however, with less than two periods over the full spectral range (Figure~\ref{Fig:FP_DriftWave_900}).

After checking and ruling out other possible causes, we confirm that the observed chromatic drift is indeed related to the etalon itself. The most probable reason is an aging of the dielectric multi-layer coatings on the etalon mirrors. However, the exact mechanism behind this could no be determined and further studies and simulations are required to figure out the root cause.

We compare our results with the findings of \citet{Terrien2021} who recently reported a chromatic drift of the Fabry-P\'erot interferometer installed at the Habitable-zone Planet Finder spectrograph. Overall, we find very similar effects, despite a rather different design of the Fabry-P\'erot devices (see Section~\ref{Sec:ComparisonTerrien}). We therefore conclude that chromatic drifts are a very common phenomenon that probably affects every Fabry-P\'erot interferometer installed at astronomical spectrographs.
This also has implications for actively stabilized Fabry-P\'erot interferometers. Here, one has to investigate if a stabilization at a single wavelength \citep[as e.g. in][]{Sturmer2017}  actually provides a significantly better wavelength calibration than no active stabilization at all.

Fortunately, as demonstrated within our study, it is possible to measure and characterize the chromatic drift of the Fabry-P\'erot interferometer using solely the standard thorium-argon spectra without the need of a special calibration source, like for example a laser frequency comb. Therefore, it also should be possible to include a correction of the Fabry-P\'erot chromatic drift in the routine wavelength calibration procedures of any high-resolution echelle spectrograph.

\begin{acknowledgements}

The authors thank the anonymous referee for helpful comments.

This research has made use of Astropy, a community-developed core Python package for Astronomy \citep{Astropy2013,Astropy2018}, and Matplotlib \citep{Hunter2007}.

This publication makes use of the Data \& Analysis Center for Exoplanets (DACE), which is a facility based at the University of Geneva (CH) dedicated to extrasolar planets data visualization, exchange and analysis. DACE is a platform of the Swiss National Centre of Competence in Research (NCCR) PlanetS, federating the Swiss expertise in Exoplanet research. 


{\color{black}TMS  acknowledgment the support from the SNF synergia grant CRSII5-193689 (BLUVES).}

{\color{black}XD received funding from the European Research Council (ERC) under the European Union’s Horizon 2020 research and innovation programme (grant agreement No 851555/SCORE).}

This work has been carried out within the framework of the National Centre of Competence in Research PlanetS supported by the Swiss National Science Foundation.

\end{acknowledgements}

\bibliographystyle{aa_url}
\bibliography{Literature}{}

\end{document}